\documentclass{article}

\usepackage{arxiv}

\usepackage[utf8]{inputenc} 
\usepackage[T1]{fontenc}    
\usepackage{hyperref}       
\usepackage{url}            
\usepackage{booktabs}       
\usepackage{amsfonts}       
\usepackage{nicefrac}       
\usepackage{microtype}      
\usepackage{lipsum}		
\usepackage{graphicx}
\usepackage{natbib}
\usepackage{doi}
\usepackage{xcolor}
\usepackage{multirow}

\title{User Interaction Analysis through Contrasting Websites Experience}

\author{{Decky Aspandi}\\
	Department of Analytic Computing\\
	University of Stuttgart\\
	Stuttgart, Germany\\
	\texttt{decky.aspandi-latif@ipvs.uni-stuttgart.de} \\
	\And
	{Sarah Doosdal} \\
	Department of Analytic Computing\\
	University of Stuttgart\\
	Stuttgart, Germany\\
	\texttt{sarah.dosdall@t-online.de} \\
	\And
	{Victor Ülger} \\
	Department of Analytic Computing\\
	University of Stuttgart\\
	Stuttgart, Germany\\
	\texttt{victor.uelger@web.de} \\
	\And
	{Lukas Gillich} \\
	Department of Analytic Computing\\
	University of Stuttgart\\
	Stuttgart, Germany\\
	\texttt{gil\_mail@gmx.de} \\
	\And
	{Steffen Staab} \\
	Department of Analytic Computing\\
	Universitat Stuttgart\\
	Stuttgart, Germany\\
	\texttt{steffen.staab@ipvs.uni-stuttgart.de} \\
}

\renewcommand{\shorttitle}{User Interaction Analysis through Contrasting Websites Experience}

\hypersetup{
pdftitle={User Interaction Analysis through Contrasting Websites Experience},
pdfsubject={Affective-Computing, Human-Computer-Interaction},
pdfauthor={Aspandi D., Kumar C., Christoph S., Staab S.},
pdfkeywords={Human Computer Interaction, Affective Computing, Eye-Gaze Analysis},
}

\begin{document}
\maketitle

\begin{abstract}
	Current advance of internet allows rapid dissemination of information, accelerating the progress on wide spectrum of society. This has been done mainly through the use of website interface with inherent unique human interactions. In this regards the usability analysis becomes a central part to improve the human interactions. However, This analysis has not yet quantitatively been evaluated through user perception during interaction, especially when dealing wide range of tasks. In this study, we perform the quantitative analysis the usability of websites based on their usage and relevance.  We do this by reporting user interactions based user subjective perceptions, eye-tracking data and facial expressions based on the collected data from two different sets of websites. In general, we found that the user interaction parameters are substantially difference across website sets, with a degree of relation with perceived user emotions during interactions. 
\end{abstract}


\section{Introduction}
Usability analysis, which is critical to improve the use of particular systems, has risen considerably~\cite{10.1115/1.4029750}. This analysis has been applied to wide spectrum of software applications, such as medicine \cite{med_usability} \cite{Rogers2005UsabilityTA}, schooling \cite{mazzoleni2008development} \cite{elearning} and other related fields \cite{10.1093/bioinformatics/btn633}. In general, the usability-tests with the use of  questionnaires and heuristic evaluations are considered to be the most effective approach \cite{7023887}. 


Unlike heuristic evaluation where a group of experts tries to find problems per discussion over usability standards, in usability tests, a group of participants perform tasks on the software possibly revealing usability problems. This approach is recently combined with eye-tracking approaches \cite{ehmke2007identifying}, which capitalises on the the eye-mind hypothesis suggesting that the gaze can reveals ones thoughts, thus potentially aids on evaluating user interests more objectively \cite{Poole05eyetracking},\cite{ehmke2007identifying}.


The general conventions of good usability in majority relies on the level of easiness of a website on their first use (learn ability), its efficiency, how memorizable it is, its error-prevention and how satisfiable it is to the user \cite{cappel2007usability}. To our knowledge, however, these usability terms are not fully utilised and applied to a wide range of tasks. While there are usability analyses on single websites, there is a lack of analyses in comparing websites with same use-cases, such as on websites for online shopping (for instance). 

Evaluating these usability terms is quite a challenging task especially for satisfaction. Questionnaires can show how the user perceived a website but will not necessarily show concrete problematic components. Think aloud protocols provide more detailed insight in the users thoughts but disturb the natural workflow. Hence create artificial stress possibly falsifying an usability study. Both can give insight in the users level of satisfaction but might not reveal subconsciously perceived usability problems  \cite{Landowska_IT_EmotionUsabilityEvaluation}.

This problem could be mitigated by involving the affect recognition, that have been shown to be relevant in other tasks, such as medicine \cite{4598899, quimFG}, school/learning \cite{Duo2012AnES,bjet_education} and other fields \cite{picard1997ective, aspandi2020latent,aspandi2021enhanced,affect_marketing}. However still to date, the adoption of this technique to usability analysis is still scarce (i.e. to use of automatic emotion recognition to improve Human Computer Interaction). Hence, this paper targets on performing an usability analysis on several website uses as well as identifying respective users responses. Furthermore we investigate the use of emotion recognition on usability analysis. Thus, our contributions in this work are: 

\begin{enumerate}
    \item We present the quantitative comparisons of user interaction information during different website experiences. 
    \item We shows that indeed the user interactions are substantially different across four different tasks.
    \item We highlight the relationship within user subjective perceptions (SUS), their interactions and attentions (Gaze) and the perceived expressed emotion (Facial based) for each different websites experience. 
    
\end{enumerate}

\section{Related Works}
Given its importance in many fields, the website usability analysis is gaining popularity. One of the early website usability analysis can be found in \cite{cappel2007usability}, where the selection of websites (500 company websites) are analysed through several aspects for quantitative and objective evaluations. Even though these works were able to analyse the websites presentation and navigation problems, the general statement about the quality of the websites are still not fully evaluated.

Thus the use of more specific tools such as Eye-tracking and the think-aloud technique were started to be incorporated, given its ability to give more targeted user attentions and perceptions \cite{weichbroth2016eye}. The common approach is to use various instructions, questionnaires and tasks that allows for common patterns and behaviours of the users could be identified. One such work is the work of Ehmke et al.~\cite{ehmke2007identifying} who uses eye-tracking data to identify usability problems. Specifically, different eye-tracking patterns are assigned to certain usability problems by involving 19 participants and two websites. Here they found common patterns of the usability problems, even though the main limitations are the narrowed tasks involved. 


Recent work of \cite{landowska2016limitations} raises other alternative solutions to improve the User Interface analysis, which highly related with usability interactions. The author suggest that the use of perceived user emotions could enhance user experience (UX), because the considerations of inherent human aspects. Furthermore, it also potentially reduces the dependency of the needs of indirect questionnaires which is highly laborious. However in this note, despite wide use of emotion recognition (ER) on other fields, such such as health~\cite{4598899}, teaching \cite{Duo2012AnES, bjet_education} etc. Its direct applications of website usability analysis is fairly limited. The only closest work is the work of~\cite{landowska2016limitations} that apply ER to software analysis, that is highly different to website interactions.

\section{Methodology}
\begin{figure}[t]
  \centering
\includegraphics[width=\columnwidth]{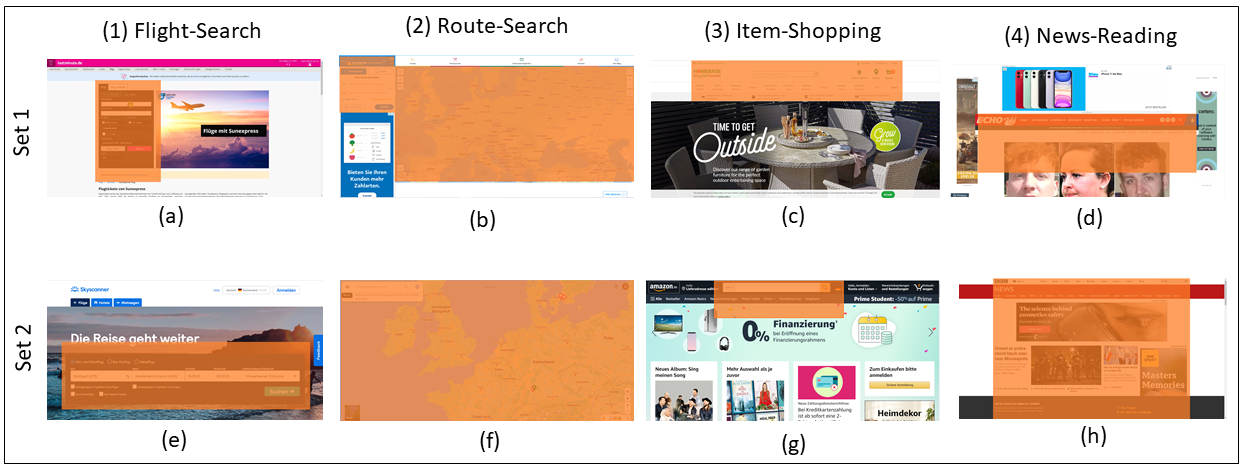}
\caption{Example of two sets of websites used in our studies, divided into four groups: Flight-Search, Route-Search, Item-Shopping and News-Reading. (a) LastMinute (b) ViaMichelin (c) HomeBase (d) LiverpoolEchos (e) SkyScanner (f) GoogleMaps (g) Amazon (h) BBC. The orange overlays show the area of interest to complete the associated tasks.}
\label{fig:websites}
\end{figure}

\subsection{Website selections}
\label{subsec:websiteSel}
We selected four most frequent tasks according to~\cite{infoplease} to assess different user interactions: 

\begin{enumerate}
\item Flight-Search: the user must find a cheapest flight from between two pre-selected locations (i.e from Stuttgart to Amsterdam).
    \item Route-Search: Similar to Flight-Search, that the user has to find the shortest route from two locations (i.e from Berlin to London). In this case however, the route solutions can be solved using different means (by foot and public transport).
    \item Item-Shopping: The user must find a predefined items (i.e in this case is both bathtub and smartwatch) which meets the users expectations.
    \item News-Reading: The user mus select an article (we chose Brexit in this case, due to the popularity).
\end{enumerate}
We selected two different website for each tasks according to their usability level. That is, the Set 1 consists of HomeBase, LiverpoolEchos, via Michelin and LastMinute websites, and Set 2 consists of Amazon, BBC, GoogleMaps and SkyScanner websites. The main differences between these two sets are that in general, Set 1 exhibits the poor usability characteristics, as explained on \cite{cappel2007usability} \cite{ehmke2007identifying}. Some examples of these characteristics are the absence of a breadcrumb trail, the use of a splash screen, overloaded presentation (thus ineffective) presentation, unclear grouping with excessive information. These poor usability aspects are persists on the Set 1 website, whereas it is considered minimal on the Set 2 (the examples of these different website designs and presentation can be seen in the Appendix~\ref{append:a}). Thus, we expect that the users interactions will be respectively poor for Set 1, compared to Set 2 (as we will show on the experiment results detailing these differences). Finally, the visualization of each website sets can be seen in the Figure~\ref{fig:websites}. Here we also show the area of interest required to complete the tasks (i.e the User Interface area the participant need to engage-thus focus- to be able to complete the desired objectives).

\begin{figure}[t]
  \centering
\includegraphics[width=170px]{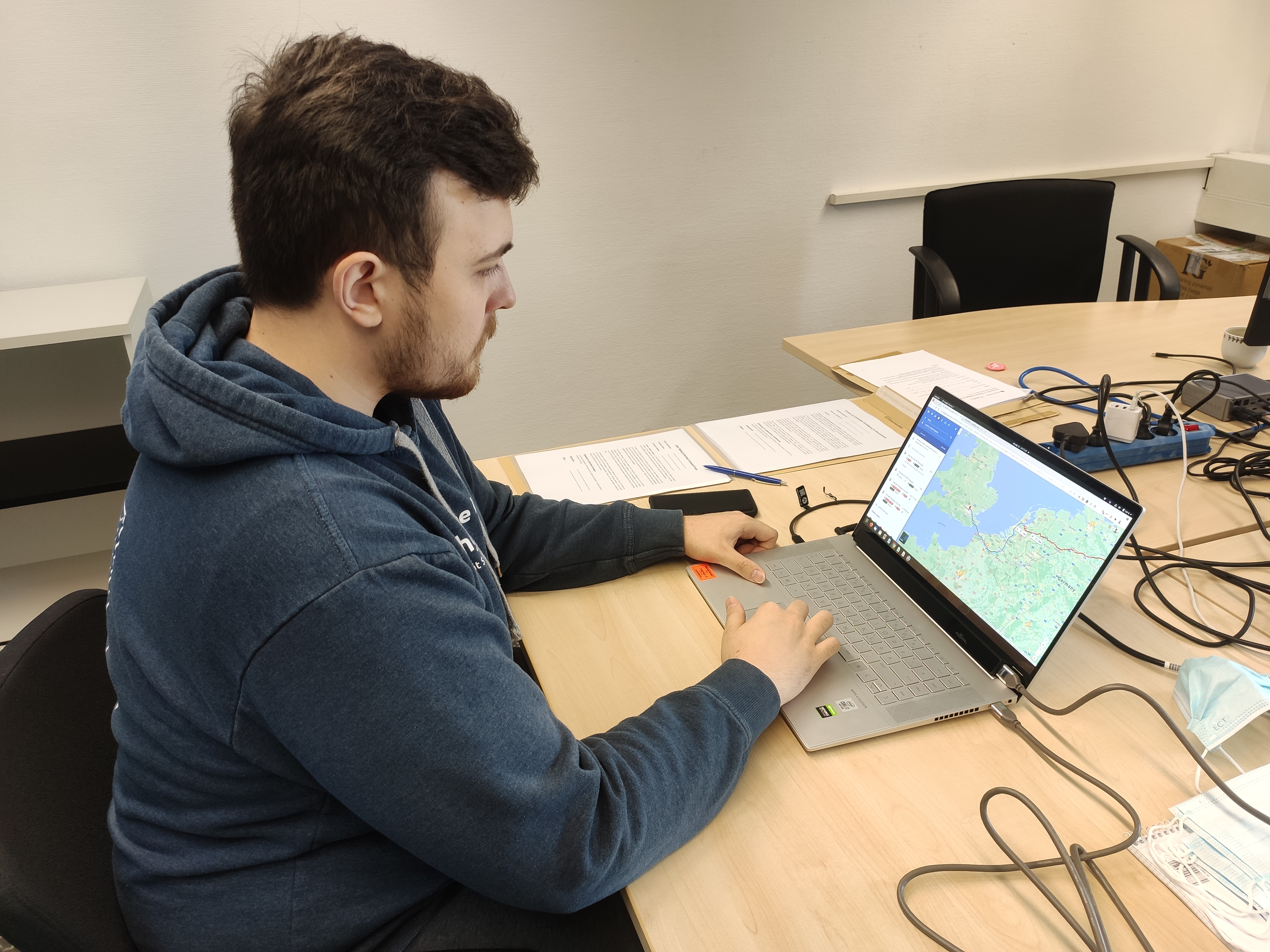}\\

  \caption{Example of data gathering process with a laptop with the connected eyetracker and webcam.}
  \label{fig:pipeline}
\end{figure}

\subsection{Data gathering}
We recorded three main modalities from the participants: The SUS score questions~\cite{brooke1996sus}, User Intractions including Gaze Data, and Facial area. The SUS score provides users subjective opinion that consists of 10 questions identifying the overall user perceptions (the example of SUS score can be seen in the Appendix~\ref{append:b}). User interactions and Gaze data provides the overview of the interactions and visual focus of the users during interacting with the websites. The user interaction consists of session duration, mouse clicks and number of pages, while Gaze data includes the Gaze Scanpath and Heatmaps. Finally, the Facial area used to infer the user emotion states during the interactions. We collected the first two modalities using EyeVido platforms, while the Facial area is recording with in-house software. Additionally, we collected the participants information, such as age, gender etc. The program used to record and process the data can be found on our repository \footnote{https://github.com/deckyal/UsabilityContrastive}.

The recording was done on two separate sessions, that was alternatively rotated between sets to minimize the learning effect. Thus half of participants commenced the recording with Set 1, and later interacting with Set 2. While the other half proceed in opposite way (Set 2, then Set 1). Prior to the recording, the users acceptance forms were signed and then they were seated in front of laptop equipped with eye-tracker and web-cams to start the recordings (the example of recording setting can be seen in the Figure~\ref{fig:pipeline}). The recording were performed in Universitat Stuttgart and took about three months to complete. In total, we successfully recorded 16 participants (11 male, 5 female) with mean age of 25 and a standard deviation of 8.

\begin{figure}[t!]
 \centering
\includegraphics[width=0.55\linewidth]{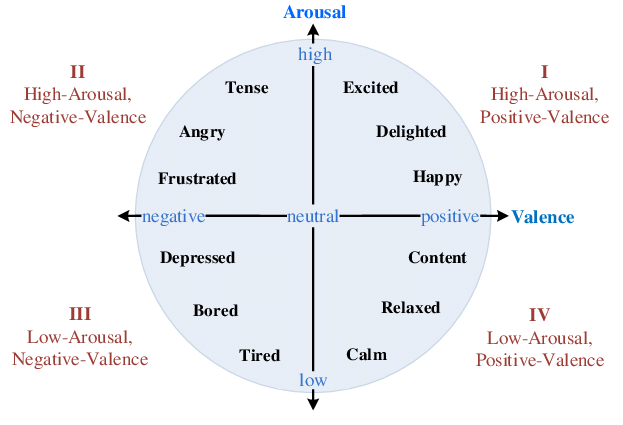}\\
\caption{Example of the Valence and Arousal dimensional space (taken from  \cite{yu2016building}).}
 \label{fig:va_space}
\end{figure}

\subsection{Data processing}

Given recorded datasets, we process three modalities separately to allow for specific analysis: 
We first harmonize the SUS results by removing the alternative question scheme resulting to positive connotations. This is done inverting the score acquired from the odd numbered questions, which also imply changing the connotation for analysis (from negative to positive). For instance, applying this invertion to SUS 2, will change the questions to be "I do not found the website in this set unnecessarily complex". This allow us to provide consistent interpretations, as the higher the score means the more positve user perceptions. Additionally, we also convert the resulting score to the maximum of 0-100. Furthermore, we directly compute each constituent of User Interactions and Eye-Gaze directly Given wide array of functionality that EyeVido has.

We use facial area to extracts the user emotion identity by using two different emotion representations: Discrete emotion identity~\cite{ekman1992there}: (Angry, Disgusted, Fearfull, Happy, Netural, Sad and Surprised) and Valence and Arousal (VA)~\cite{yu2016building} (the visualization can be seen on the Figure~\ref{fig:va_space}). The discrete emotion identity provides directly interpretable and rigid user emotions, while VA provides wider emotion examples.  Specifically, the Valence provides the positive level of user emotions while Arousal provides the corresponding activity. Thus high level of both Valence and Arousal suggests negative emotions of the user (Qaudrant I), while the low value of Valence means strong negative emotions are present (Quadrant II).

We first detect the facial area from the video using pretrained Multi-task Cascaded Convolutional Network (MTCNN)~\cite{DBLP:journals/corr/ZhangZL016}. Then given the located facial area, we apply deep learning based (Convolutional Recurrent Neural Networks) VA estimations model~\cite{kolliasAffWildNet} which produces the respective VA identify (ranged between 0 to 1 continuous values). Then, we further utilize similar logistic based Convolutional neural network based model \footnote{https://github.com/atulapra/Emotion-detection} to infer the discrete emotion labels.

\section{Results and Analysis}

\subsection{Sus score analysis}

\label{sec:SUSAnalysis}
\begin{figure}[t]
  \centering
\includegraphics[width=\columnwidth]{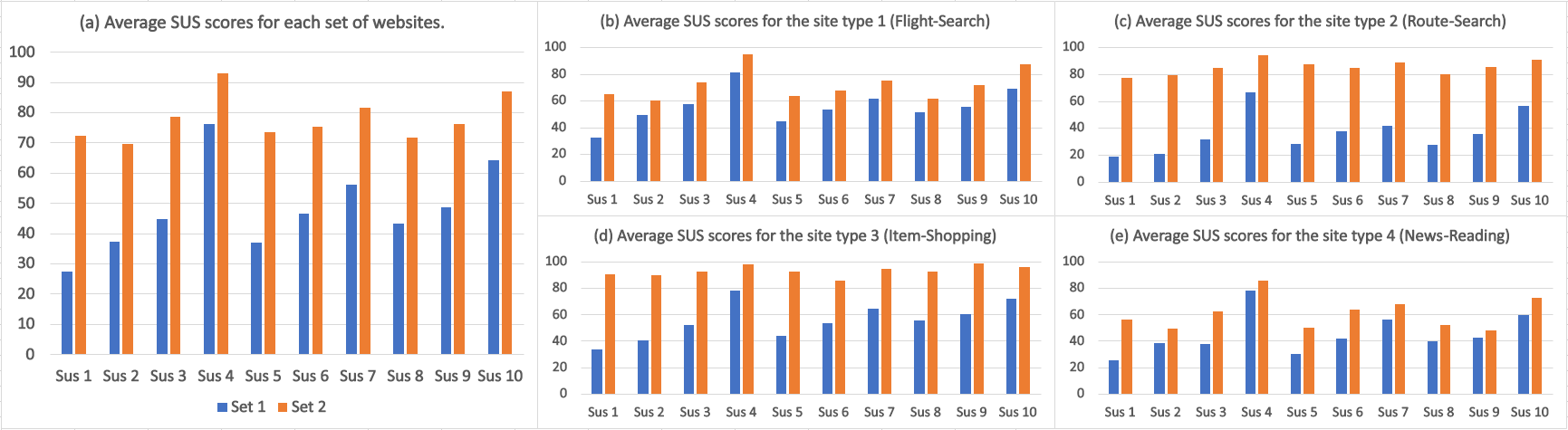}

  \caption{SUS score evaluations: a). The average SUS scores for each set/question; b) The SUS scores for each type 1 website (Flight-Search); c) The SUS scores for each type 2 website (Route-Search); d) The SUS scores for each type 3 website (Item-Shopping); e) The SUS scores for each type 4 website (News-Reading).   }
  \label{fig:sus}
\end{figure}

\begin{table}[t]
\caption{The average SUS score results for each set and every question, together with the average differences between the sets and the average scores for each question for all websites combined.}
\begin{tabular}{llllllllllll}
\hline
\hline
Websites & Sus 1 & Sus 2 & Sus 3 & Sus 4 & Sus 5 & Sus 6 & Sus 7 & Sus 8 & Sus 9 & Sus 10 & AVG \\ \hline
Set 1 & 27,3 & 37,2 & 44,5 & 76,0 & 36,7 & 46,4 & 56,0 & 43,2 & 48,4 & 64,1 & 48,0 \\ \hline
Set 2 & 72,1 & 69,5 & 78,4 & 93,0 & 73,4 & 75,3 & 81,5 & 71,6 & 76,0 & 86,7 & 77,8 \\ \hline
\textbf{AVG differences} & \textbf{44,8} & \textbf{32,3} & \textbf{33,9} & \textbf{16,9} & \textbf{36,7} & \textbf{28,9} & \textbf{25,5} & \textbf{28,4} & \textbf{27,6} & \textbf{22,7} & \textbf{29,8} \\ \hline
\hline
\end{tabular}%
\label{tab:sus1}
\end{table}

\begin{table}[t]
\caption{Overview of all individual SUS scores for each website/question combination. This also includes the difference between the individual website types between the sets. Red scores indicate the highest differences between the sets for a single site type, while blue scores indicate the lowest counterparts}
\resizebox{\textwidth}{!}{%
\begin{tabular}{llllllllllll}
\hline
\hline
 Websites & Sus 1 & Sus 2 & Sus 3 & Sus 4 & Sus 5 & Sus 6 & Sus 7 & Sus 8 & Sus 9 & Sus 10 & AVG \\ \hline
Flight-Search Set1 & 32,3 & 49,0 & 57,3 & 81,3 & 44,8 & 53,1 & 61,5 & 51,0 & 55,2 & 68,8 & 55,4 \\ \hline
Flight-Search Set2 & 64,6 & 60,4 & 74,0 & 94,8 & 63,5 & 67,7 & 75,0 & 61,5 & 71,9 & 87,5 & 72,1 \\ \hline
\textbf{AVG differences} & \textbf{32,3} & \textbf{11,5} & \textcolor{blue}{\textbf{16,7}} & \textbf{13,5} & \textcolor{blue}{\textbf{18,8}} & \textcolor{blue}{\textbf{14,6}} & \textbf{13,5} & \textcolor{blue}{\textbf{10,4}} & \textbf{16,7} & \textbf{18,8} & \textbf{16,7} \\ \hline
\hline
Route-Search Set1 & 18,8 & 20,8 & 31,3 & 66,7 & 28,1 & 37,5 & 41,7 & 27,1 & 35,4 & 56,3 & 36,4 \\ \hline
Route-Search Set2 & 77,1 & 79,2 & 84,4 & 93,8 & 87,5 & 84,4 & 88,5 & 80,2 & 85,4 & 90,6 & 85,1 \\ \hline
\textbf{AVG differences} & \textcolor{red}{\textbf{58,3}} & \textcolor{red}{\textbf{58,3}} & \textcolor{red}{\textbf{53,1}} & \textcolor{red}{\textbf{27,1}} & \textcolor{red}{\textbf{59,4}} & \textcolor{red}{\textbf{46,9}} & \textcolor{red}{\textbf{46,9}} & \textcolor{red}{\textbf{53,1}} & \textcolor{red}{\textbf{50,0}} & \textcolor{red}{\textbf{34,4}} & \textcolor{red}{\textbf{48,8}} \\ \hline
\hline
Item-Shopping Set1 & 33,3 & 40,6 & 52,1 & 78,1 & 43,8 & 53,1 & 64,6 & 55,2 & 60,4 & 71,9 & 55,3 \\ \hline
Item-Shopping Set2 & 90,6 & 89,6 & 92,7 & 97,9 & 92,7 & 85,4 & 94,8 & 92,7 & 99,0 & 95,8 & 93,1 \\ \hline
\textbf{AVG differences} & \textbf{57,3} & \textbf{49,0} & \textbf{40,6} & \textbf{19,8} & \textbf{49,0} & \textbf{32,3} & \textbf{30,2} & \textbf{37,5} & \textbf{38,5} & \textbf{24,0} & \textbf{37,8} \\ \hline
\hline
News-Reading Set1 & 25,0 & 38,5 & 37,5 & 78,1 & 30,2 & 41,7 & 56,3 & 39,6 & 42,7 & 59,4 & 44,9 \\ \hline
News-Reading Set2 & 56,3 & 49,0 & 62,5 & 85,4 & 50,0 & 63,5 & 67,7 & 52,1 & 47,9 & 72,9 & 60,7 \\ \hline
\textbf{AVG differences} & \textcolor{blue}{\textbf{31,3}} & \textcolor{blue}{\textbf{10,4}} & \textbf{25,0} & \textcolor{blue}{\textbf{7,3}} & \textbf{19,8} & \textbf{21,9} & \textcolor{blue}{\textbf{11,5}} & \textbf{12,5} & \textcolor{blue}{\textbf{5,2}} & \textcolor{blue}{\textbf{13,5}} & \textcolor{blue}{\textbf{15,8}} \\ \hline
\hline
\end{tabular}%
}
\label{tab:sus2}
\end{table}



Table~\ref{tab:sus1} shows the overall SUS score of all websites between Set 1 and Set 2, with Table~\ref{tab:sus2} provides the score for each websites. The Figure~ \label{fig:sus} also shows corresponding graph. Here we can see that in overall, the Set 1 scores almost 30\% (29.8) less than Set 2 indicating the strong user preferences to the Set 2. The SUS 1 ("I would like to use this website frequently") in particular shows the most margin between set, which is due to the global notion of the questions asked. This in contrast to SUS 4 ("I think that I would not need the support of a technical person to be able to use this system") where the margin is the smaller. Given this information,  we can say that in general, people tend to use their preferred website frequently. However in either case, the technical supports are still deemed necessary.

Specific to the SUS score for each website, we found that the biggest SUS score margin on the route search, with the minimum margin on the News-Reading task. This implies that the differences of the website designs on route search are more pronounced compared to news reading. One such examples are the huge area of ads that the ViaMicheline has, while it is not exists on GoogleMaps (this further also impacts the user attentions, as shown on the next sections). As for the overall SUS score, we found that the item shopping task yields the highest score (93.1, Set 2) which suggests that the user are quite comfortable with current website designs for shopping. Whereas, the lowest SUS score is shown the route-search task (36.4, Set 1) indicating user dislike on current website design for the route search (suggesting current limitation of such designs).

\subsection{User Interaction and Eye Gaze parameter analysis}
\subsubsection{User Interaction Analysis}
\label{subsec:userInteract}

\begin{figure}[t!]
  \centering
\includegraphics[width=\columnwidth]{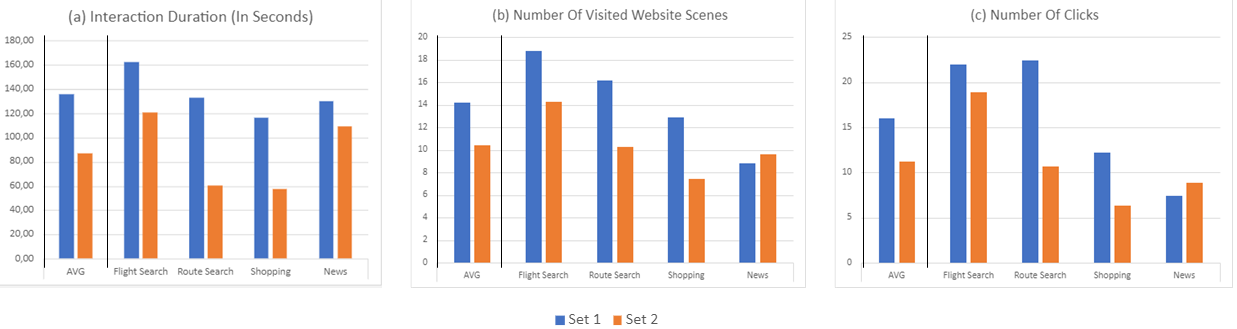}\\
\caption{The graphs of collected interactions data: (a) the duration of the interactions (seconds); (b) how many scenes were displayed per web page; (c) the amount of clicks that have been made.}
  \label{fig:interactionanalysis}
\end{figure}

\begin{table}[]
\centering
\begin{tabular}{l|ccccc}
\hline
\hline
Interaction Parameters      & Flight-Search & Route-Search & Shopping & News  & \textbf{AVG}   \\
\hline
\hline

 Interaction Duration (in Second) - Set 1 & 162,4         & 132,8        & 116,7    & 130,0 & \textbf{135,5} \\ \hline
Interaction Duration (in Second) - Set 2 & 120,8         & 60,0         & 57,31    & 109,3 & \textbf{86,9}  \\ \hline \hline

Number Of Visited Website Scenes - Set 1 & 21,9& 22,4         & 12,2     & 7,4   & \textbf{15,9}  \\ \hline
Number Of Visited Website Scenes - Set 2 & 18,9& 10,7         & 6,3      & 8,9   & \textbf{11,2}       \\ \hline \hline
Number Of Clicks - Set 1 & 18,8& 16,2         & 12,9     & 8,8   & \textbf{14,1}  \\ \hline
Number Of Clicks - Set 2 & 14,3& 10,3         & 7,4      & 9,6   & \textbf{10,4}  \\ \hline
\hline
\end{tabular}

\caption{The table shows the corresponding values in Figure \ref{fig:interactionanalysis}. The first two rows are duration of the visit to the website in seconds. The second and third rows are the number of viewed scenes. Last two rows are the average number of clicks made by the users.}
\label{tab:UserInteractionData}
\end{table}

The Table~\ref{tab:UserInteractionData} shows the interactions data for all websites with corresponding graph shown in the Figure~\ref{fig:interactionanalysis}. We can observe that in this table, the Set 2 required about 30\% less interaction (duration, scenes and number of clicks) compared to Set 1. This could be explain from our finding in SUS analysis which shows user preference toward the second Set (SUS 1 as the highest margin). Thus, this suggests that the less interaction, in general, corresponds to the higher user perceptions (i.e more efficient). This is further conformed by the highest difference in the Route-Search task across categories (with almost twice the values)



 \subsubsection{Eye gaze Parameter}
 \label{sec:gazeparameter}

\begin{figure}[t!]
  \centering
\includegraphics[width=\columnwidth]{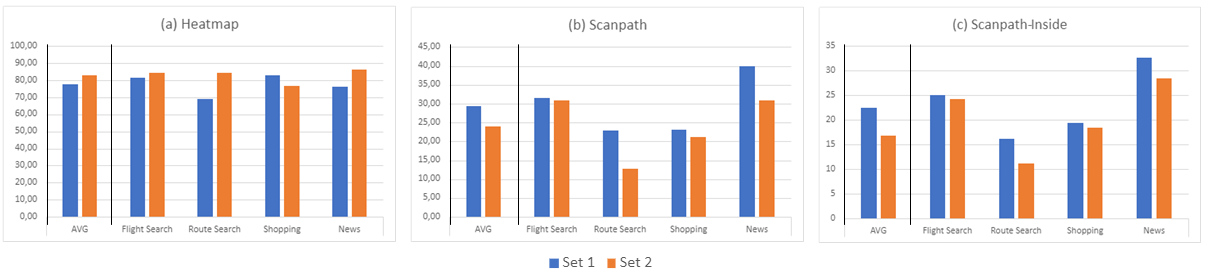}\\

\caption{Graph of collected eye-gaze data. (a)Heatmap - shows the percentage of fixation in the area of interest. (b) Average length of scanpath, (c) Number of scanpath points that were within the area of interest.}
  \label{fig:eyegaze}
\end{figure}

\begin{table}[]
\centering
\begin{tabular}{l|ccccc}
\hline
\hline
Interaction Parameters         & Flight-Search & Route-Search & Shopping & News & \textbf{AVG}  \\ \hline

Heatmap - Set 1    & 81,6& 68,9         & 83,1     & 76,2 & \textbf{77,4} \\ \hline
Heatmap - Set 2    & 84,2& 84,2         & 76,9     & 86,2 & \textbf{82,8} \\ \hline  \hline

Scanpath - Set 1    & 31,5& 22,9         & 23,1     & 39,9 & \textbf{29,3} \\ \hline
Scanpath - Set 2    & 30,8& 12,7         & 21,2     & 30,8 & \textbf{23,9} \\ \hline
 \hline
Scanpath-Inside - Set 1    & 24,9& 16,1         & 19,3     & 32,6 & \textbf{22,0} \\ \hline
Scanpath-Inside - Set 2    & 24,2& 11,1         & 18,3     & 28,4 & \textbf{17,0} \\ \hline
\hline
\end{tabular}
\caption{This table shows the associated values from Figure \ref{fig:eyegaze}. (a) Heatmap - represents the percentage of fixation in the area of interest. (b)Full length scanpath, (c)Scanpath - inside, are the parts of the scanpath that were inside the area of interest.}
\label{tab:EyeGazeData}
\end{table}

\begin{figure}[t!]
  \centering
\includegraphics[width=\columnwidth]{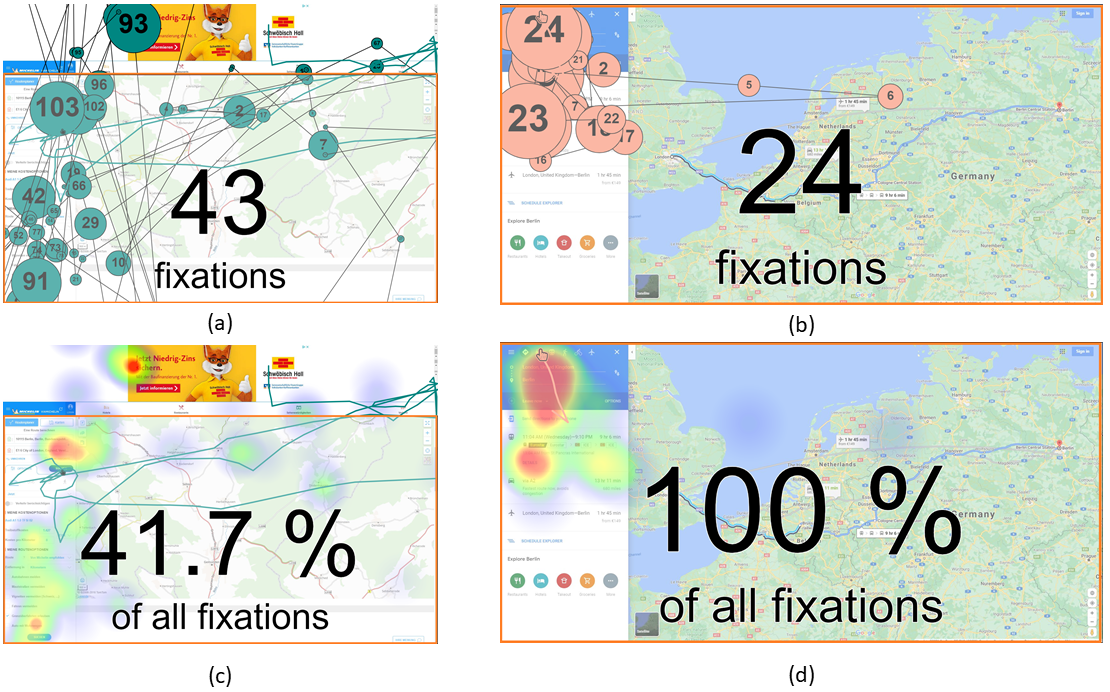}\\
\caption{Example data gathering process (a) SCP inside is 43 out of 103 total of SCP. (b) SCP inside is 24 out of 24 total of SCP. (c) Percentage of Heatmap inside is 41.7\%. (d) Percentage of Heatmap inside is 100\%.}
  \label{fig:areaofinterest}
\end{figure}

The Table~\ref{tab:EyeGazeData} shows the extracted Eye-gaze parameters (Heatmaps, overall scanpath, and scan path inside area of interest-Scannpath Inside) with respective graph visualized in Figure~\ref{fig:eyegaze}. We can see that in overall the values of heatmaps are larger, with lower scanpath values for Set 2 compared to the Set 1. This means, the user focus more in the relevant area (intensive heatmaps on the shaded orange, cf Section ~\ref{subsec:websiteSel}) with less visual interference (low scan-path values). The examples of the heatmaps and fixations from both sets for route-search are shown on the Figure~\ref{fig:areaofinterest} (we chose route-search given its largest margin of SUS score, thus allowing for more pronounced differences). Here we see that within the are of interests (on orange box), the heatmaps percentage are larger for Set 1 with lower fixations level. Furthermore, we can also see some major distractions such as ads are shown, that distracts the user focus (thus unnecessary gaze instances-higher scanpath).




\label{fig:record}

\subsection{Facial emotion based analysis}

\subsubsection{Valence and Arousal based Emotion Analysis} 

\begin{figure}[t!]
    \centering
    \includegraphics[width=\columnwidth]{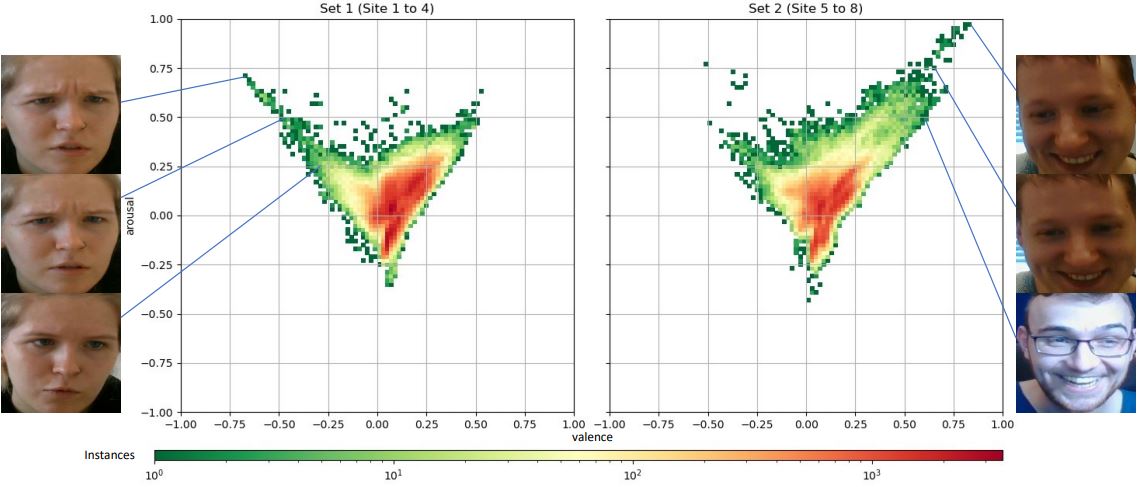}\\
    \caption{Distribution of our measurements in a 2D Valence Arousal space for evaluated websites, alongside the examples of facial area. Left: the results on website 1 to 4 (Set 1). Right: the results on website 5 to 8 (Set 2).}
    \label{fig:vaHist2d}
\end{figure}

Figure~\ref{fig:vaHist2d} shows the distributions of predicted VA estimations of the participants on both Sets (Set 1 to the left, with Set 2 on the right) with associated facial area examples. We can see that in overall, there is a tendency of the VA distribution of Set 1 to the second quadrant (negative valence, and positive arousal) suggesting strong negative emotion states (such as angry in extreme end of spectrum).  While the VA distribution of Set 2 tend to distribute on the first quadrant (positive valence and arousal), which shows more positive emotion perceived (such as Delighted). These different distribution patterns, not only shows the strong different emotion states between sets, but also in line with our previous results on both subjective user perspective (SUS) along with their more efficient task completions(user interaction and gaze). 



Since VA based emotion recognition allows for continuous prediction over time, then we extracted these values during the users completion of respective tasks. In this Figure, we can observe that the dynamics are more profound on the Set 2 compared to the Set 1. This is indicated by the constant increase of arousal identity during the interaction with website on Set 2, while it is fairly constant on the Set 1. This suggests that the users are more engaged on Set 2 than the Set 1. Furthermore, this is also shown on the associated statistical values on the Table~\ref{tab:vaMeasures}, that in overall, the mean and median valence values are higher on the Set 2 compared to Set 1. 




\begin{figure}[t!]
    \centering
    \includegraphics[width=\columnwidth]{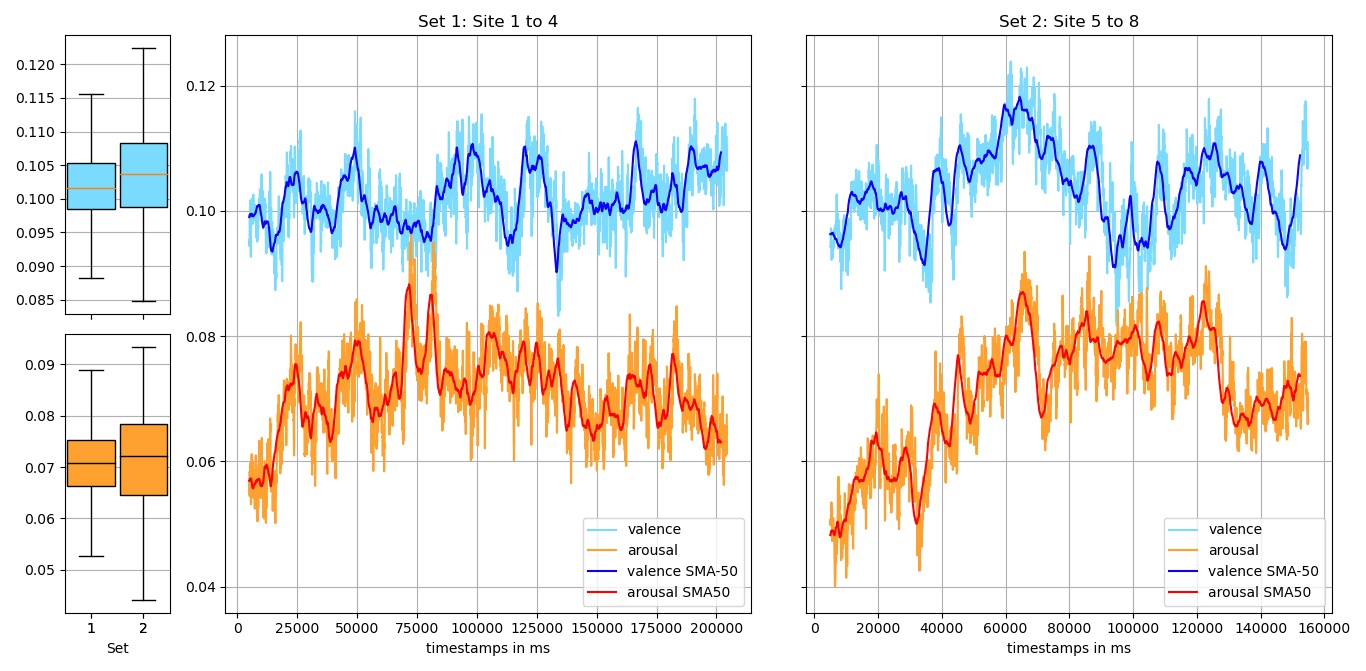}\\
    \caption{Valence and Arousal annotations averaged over the 16 users, illustrated as (left) boxplot, (middle) progress at Set 1 over time (averaged among sites 1 to 4) and (right) progress at Set 2 over time (averaged among sites 5 to 8).}
    \label{fig:vaboxtime}
\end{figure}

\begin{table}[]
\centering
\caption{Valence and Arousal values for all users on all evaluated websites. The left column shows the mean results for each websites. The right column shows the statistical values for all websites on the respective set. }
\begin{tabular}{c|cccc|ccccc}
\hline
\hline
\multirow{2}{*}{Set 1} & \multicolumn{4}{c|}{All Users} & \multicolumn{5}{c}{\textbf{All Websites}} \\ \cline{2-10} 
 & Website 1 & Website 2 & Website 3 & Website 4 & \textbf{Mean} & \textbf{Median} & \textbf{STD} & \textbf{Min} & \textbf{Max} \\ \hline
Valence & 0,102 & 0,099 & 0,101 & 0,106 & 0,102 & 0,102 & 0,005 & 0,083 & 0,118 \\ \hline
Arousal & 0,067 & 0,081 & 0,065 & 0,070 & 0,071 & 0,071 & 0,007 & 0,050 & 0,097 \\ \hline
\hline
\multirow{2}{*}{Set 2} & \multicolumn{4}{c|}{All Users} & \multicolumn{5}{c}{\textbf{All Websites}} \\ \cline{2-10} 
 & Website 5 & Website 6 & Website 7 & Website 8 & \textbf{Mean} & \textbf{Median} & \textbf{STD} & \textbf{Min} & \textbf{Max} \\ \hline
Valence & 0,108 & 0,102 & 0,105 & 0,099 & 0,104 & 0,104 & 0,007 & 0,082 & 0,124 \\ \hline
Arousal & 0,074 & 0,080 & 0,061 & 0,067 & 0,071 & 0,072 & 0,010 & 0,040 & 0,093 \\ \hline
\hline
\end{tabular}%
\label{tab:vaMeasures}
\end{table}

\subsubsection{Overall Discrete emotion identity}
\begin{figure}[t!]
    \centering
    \includegraphics[width=0.65\columnwidth, trim= 0cm 0.5cm 1cm 1cm, clip]{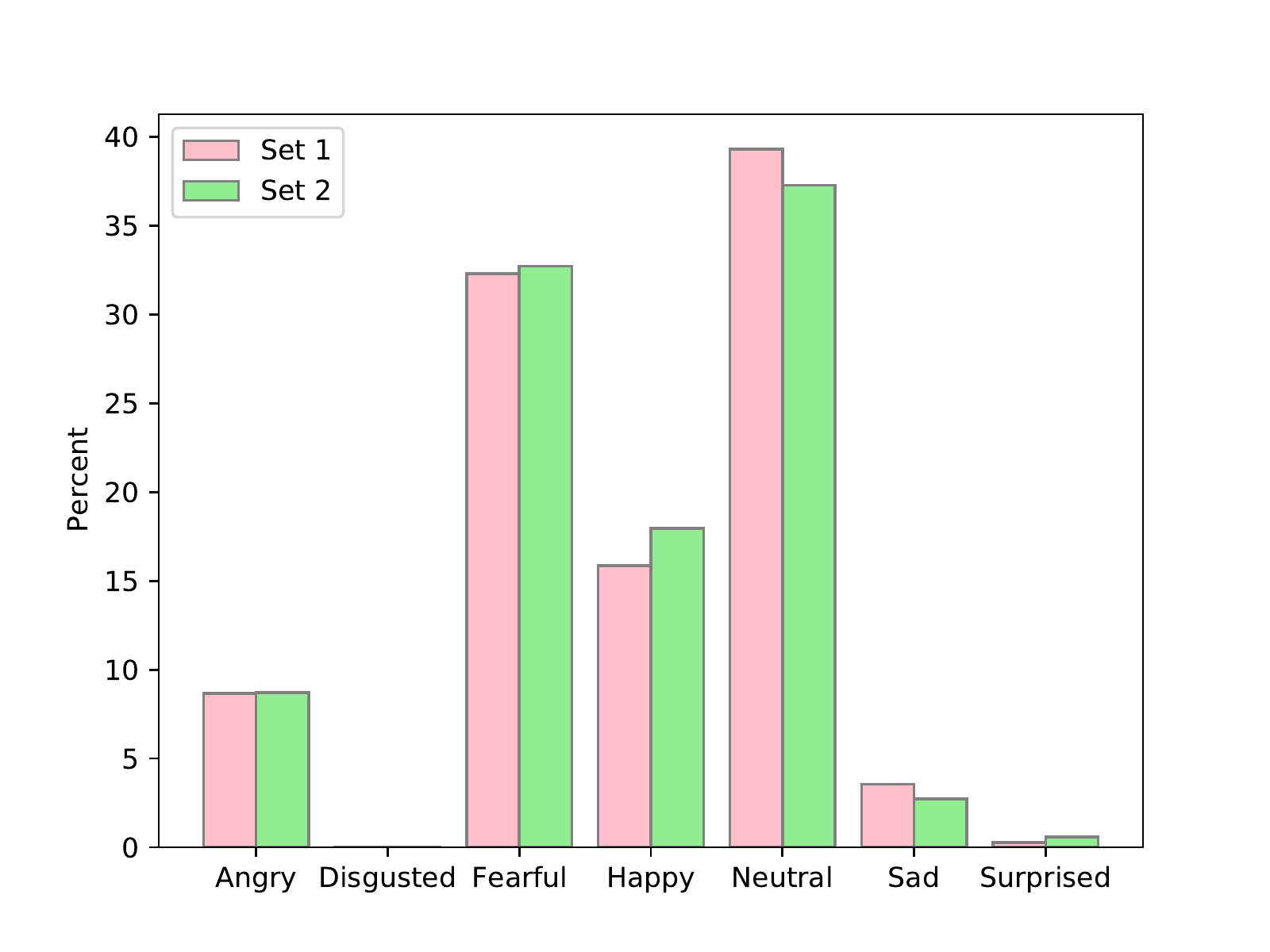}\\
    \caption{Histogram of the accumulated discrete emotion labels between Set 1 and Set 2.}
    \label{fig:discreteEmot}
\end{figure}

Figure~\ref{fig:discreteEmot} shows the graph of accumulated discrete emotion identity of all websites and users for each set. Here we see a similar results with the finding from VA based emotion analysis. That is, we observe that the predicted emotion on Set 2 tend to be positive (i.e Happy) and followed by less accumulated Sad emotion instance compared to Set 1. This highlights the general agreements of these independent emotion model suggesting the more positive emotions perceived of the user, when interacting with website on the Set 2. 


\section{Conclusions}
In this work, we evaluated user interactions on performing different tasks through distinct website sets. we do this by first defining the four most routine internet tasks: flight search, route search, item shopping and news reading. Then, we search and select associated website that contains major usability issues as Set 1, and the less counterparts as Set 2. Using our custom build software in conjunction with EyeVido platform, we record three main modalities from 16 individuals, namely System Usability Score, User Interaction and Gaze data, and Facial Based Emotion identity.  acquired the required recording consisting of 16 individuals. 

We show in our analysis that there is different user interactions between sets. In our SUS score analysis, we highlight the strong preference of user to the Set 2, with route-search task showing the substantial score difference to Set 1. Furthermore, we observe more efficient interactions and focused user attentions in our gaze and user data analysis. Which relates the positive user perception with the more efficient interactions. Finally, the extracted Valence/Arousal with Categorical emotion identity suggests more positive emotion states in the interaction of user on Set 1 compared to Set 2. Which further shows that indeed, in some degree, the user express their emotion visually during their interactions. 

In the future, the recorded datasets alongside the findings will potentially relevant to create an automatic usability predictor. Which will be relevant to advance current usability analysis, which is considered still largely manual.

\section{Acknowledgements}
The authors would like to thanks Raphael, Chandan, Ramin from Department of Analytic Computing of University of Stuttgart for the supports on data processing. The authors would further express appreciation to Christoph and Tina from EYEVIDO GmbH for the supports on data gathering. This research is supported by the funding from UDeco project by Germany BMBF KMU Innovativ.

\appendix
\section{Appendix: Website Set Selections}
\label{append:a}
Figure \ref{fig:app-data}-\ref{fig:app-data4} below show the examples of different characteristics of evaluated website sets, as explained on the Section \ref{subsec:websiteSel}. Figure \ref{fig:app-data} shows that the breadcrumbs trail is missing on the Set 1 (a). Figure \ref{fig:app-data2} visualizes the example of overloaded, thus ineffective website presentations due to irrelevant information (such as in this example, on (a)). Figure \ref{fig:app-data3} shows the example where the search results from the websites does not refer the proper product page. Lastly, Figure \ref{fig:app-data4} gives example where the search facility is lacking on the evaluated LiverpoolEcho website (Set 1); that in contrast exists on the BBC (Set 2).

\begin{figure}[t]
    \centering
      \includegraphics[width=\linewidth]{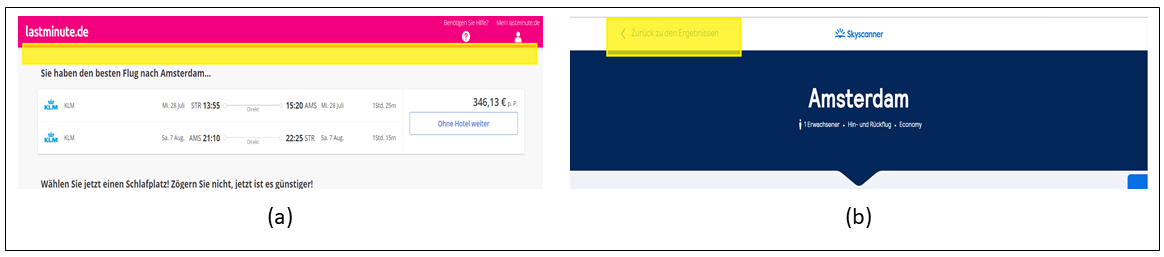}
      \caption{Flight-Search. (a) lastminute: shows absence of a breadcrumb trail (in yellow). (b) SkyScanner: shows presence of a breadcrumb trail.}
      \label{fig:app-data}
\end{figure}

\begin{figure}[t]
    \centering
      \includegraphics[width=\linewidth]{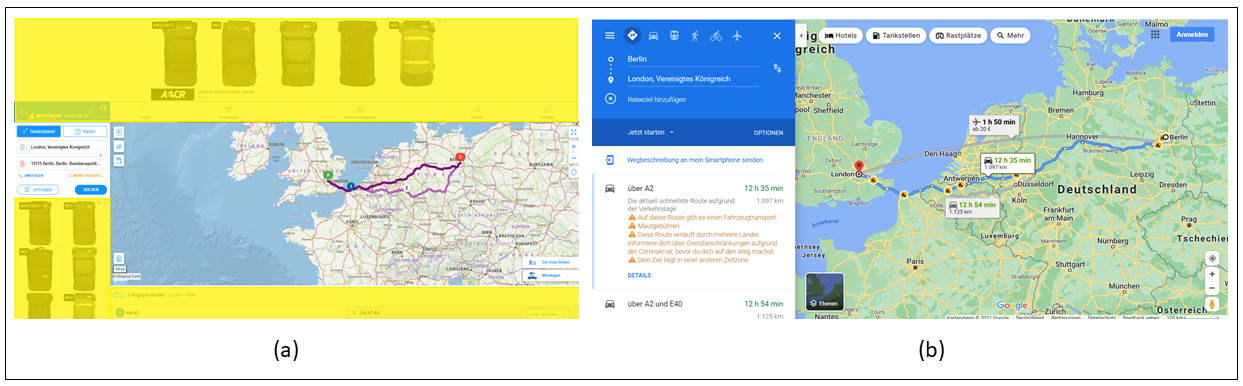}
      \caption{Route-Search: (a) ViaMicheline: has a overloaded, ineffective presentation due to ads (in yellow). (b) GoogleMaps: clear presentation of the results.}
      \label{fig:app-data2}
\end{figure}

\begin{figure}[t]
    \centering
      \includegraphics[width=\linewidth]{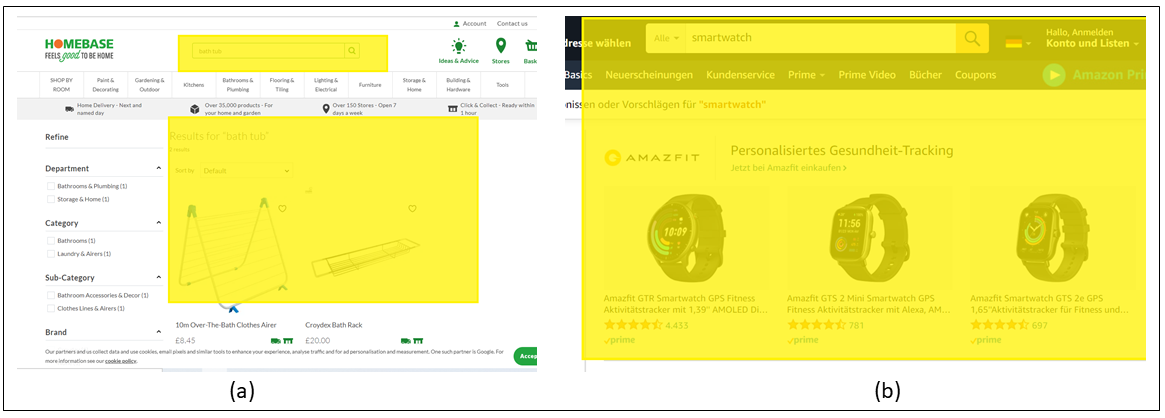}
      \caption{Item-Shopping: (a) HomeBase :Expected information is missing, searching with keyword 'bath tub' does not show bath-tub product page to the user (in yellow). (b) amazon: Searching with the keyword properly guide users to the relevant products (smartwatch in this example).}
      \label{fig:app-data3}
\end{figure}

\begin{figure}[t]
    \centering
      \includegraphics[width=\linewidth]{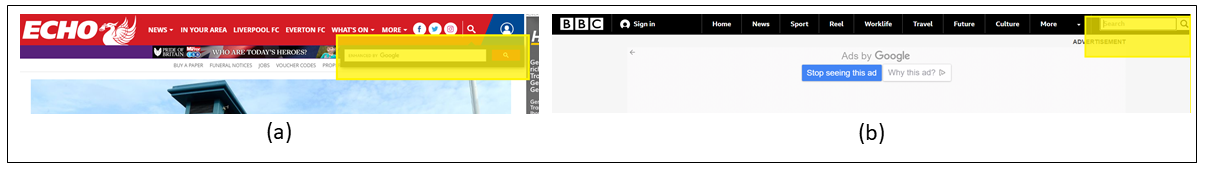}
      \caption{News-Reading: (a) LiverpoolEcho: A site search capability is not provided (in yellow). (b) BBC: A site search facility is provided.}
      \label{fig:app-data4}
\end{figure}

\section{Appendix: SUS Score}
\label{append:b}
The SUS questions used in this work are the following (scale of 0 to 7, with 7 implies strong agreement):
\\
\noindent\textbf{SUS 1.} I think that I would like to use the websites in this set frequently.\\
\textbf{SUS 2.} I found the websites in this set unnecessarily complex.\\
\textbf{SUS 3.} I thought the websites in this set were easy to use.\\
\textbf{SUS 4.} I think that I would need the support of an expert to be able to use the websites in this set.\\
\textbf{SUS 5.} I found the various functions on the websites in this set were well integrated.\\
\textbf{SUS 6.} I thought there was too much inconsistency on the websites in this set.\\
\textbf{SUS 7.} I would imagine that most people would learn to use the websites in this set very quickly.\\
\textbf{SUS 8.} I found the websites in this set very cumbersome to use.\\
\textbf{SUS 9.} I felt very confident using the websites in this set.\\
\textbf{SUS 10.} I needed to learn a lot of things before I could get going with the websites in this set.\\\\

\bibliographystyle{unsrtnat}
\bibliography{references}  






\end{document}